\documentclass[aps,prl,superscriptaddress,twocolumn,floatfix,a4paper]{revtex4}

\usepackage{graphicx,graphics,epsfig}   
\usepackage{dcolumn}    
\usepackage{bm}         
\usepackage{amsmath}    
\usepackage{verbatim}   
\usepackage{color}      
\usepackage{subfigure}  
\usepackage{times,natbib}
\usepackage{amsmath,amsfonts,amssymb,graphics,graphics,color,times}

\usepackage{latexsym}
\usepackage{amsmath}
\usepackage{amssymb}
\usepackage{amsfonts}
\usepackage{amsthm}
\usepackage{mathrsfs}
\usepackage{color,verbatim,graphics}
\usepackage{psfrag}
\DeclareMathAlphabet{\mathrsfs}{U}{rsfs}{m}{n}
\DeclareMathAlphabet{\mathpzc}{OT1}{pzc}{m}{it}
\DeclareMathAlphabet{\matheus}{U}{eus}{m}{n}
\DeclareMathAlphabet{\mathbbold}{U}{bbold}{m}{n}

\newcommand{\ba}{\begin{eqnarray}}
\newcommand{\ea}{\end{eqnarray}}
\newcommand{\ban}{\begin{eqnarray*}}
\newcommand{\ean}{\end{eqnarray*}}

\newcommand{\ket}[1]{|#1\rangle}
\newcommand{\bra}[1]{\langle#1|}
\newcommand{\braket}[2]{\langle#1|#2\rangle}

\begin{document}

\title{Quantum nonlocality does not imply entanglement distillability}

\author{Tam\'as V\'ertesi}
\affiliation{Institute of Nuclear Research of the Hungarian Academy of Sciences
H-4001 Debrecen, P.O. Box 51, Hungary}
\author{Nicolas Brunner}
\affiliation{H.H. Wills Physics Laboratory, University of Bristol, Tyndall Avenue, Bristol, BS8 1TL, United Kingdom}



\begin{abstract}
Entanglement and nonlocality are both fundamental aspects of quantum theory, and play a prominent role in quantum information science. The exact relation between entanglement and nonlocality is however still poorly understood. Here we make progress in this direction by showing that, contrary to what previous work suggested, quantum nonlocality does not imply entanglement distillability. Specifically, we present analytically a 3-qubit entangled state that is separable along any bipartition. This implies that no bipartite entanglement can be distilled from this state, which is thus fully bound entangled. Then we show that this state nevertheless violates a Bell inequality. Our result also disproves the multipartite version of a longstanding conjecture made by A. Peres.
\end{abstract}

\maketitle

The correlations obtained from performing local measurements on entangled quantum states cannot be explained by any classical mechanism: communication is excluded because the signal should travel faster than light. Pre-established agreement is excluded because Bell inequalities are violated \cite{bell}. Formally, this last step means that quantum correlations cannot be decomposed into local ones; for instance, considering three distant parties Alice, Bob and Charlie, sharing an entangled state $\rho$, we have that
\ba \label{local} P_\rho(abc|xyz) \neq \int  d\lambda   P(a | x\lambda)P(b | y\lambda)P(c | z\lambda) \ea
where $x,y,z$ denote the measurement settings of Alice, Bob and Charlie, $a,b,c$ the measurement outcomes, and $\lambda$ is an arbitrary shared variable. 

Determining which entangled states $\rho$ can lead to Bell inequality violation is an important but difficult question. While nonlocality turns out to be a generic feature of all entangled pure states \cite{gisin}, the situation is however much more complex for mixed states. There exist mixed entangled states which admit a local hidden variable model \cite{werner}, even for the most general type of measurements \cite{barrett}. But some of these states can nevertheless violate a Bell inequality if, prior to the local measurements, pre-processing by local operations and classical communication (LOCC) is performed \cite{popescu}. In this way, the 'hidden nonlocality' of these states is revealed. Importantly this pre-processing must be independent of the choice of local measurement settings made by the parties, as communication could otherwise be used as a nonlocal resource.

In the case where more than one copy of the states are available, and joint operations on these copies can be performed, the problem becomes intimately related to entanglement distillation \cite{peres96}. Let us recall that a bipartite entangled state is said to be distillable if, from an arbitrary number of copies, it is possible to extract pure entanglement by LOCC \cite{ED}. It thus follows that any entangled state that is distillable violates a Bell inequality asymptotically. The notion of distillability can be extended in different ways to the case of more than two parties, depending on whether or not several parties are allowed to join together. Here we adopt the most general definition, and say that a multipartite quantum state is distillable, when it is possible to extract, by LOCC, pure entanglement, on (at least) one bipartition of the parties.

There exist however entangled states which are not distillable, so-called 'bound entangled' states \cite{BE}. Thus, the phenomenon of entanglement displays a form of irreversibility \cite{irrevers}, in the sense that entanglement is necessary to produce a bound entangled state, although no pure entanglement can ever be extracted from it by LOCC. This leads naturally to the question of whether bound entangled states can also violate a Bell inequality \cite{peres99}. 

More generally, the main issue is to understand the exact relation between the concepts of (i) distillability of entanglement, and (ii) nonlocality \cite{RMP_horo}. On the one hand, distillability always implies nonlocality as explained above, thus (i) implies (ii). In this paper we will focus on the converse statement, i.e. whether (ii) implies (i). Specifically, consider a scenario involving $N$ parties, sharing an arbitrary number of copies of a quantum state $\rho$. The question is whether the fact that the state $\rho$ violates a Bell inequality always implies that the parties can distill (from an arbitrary number of copies of $\rho$) some bipartite entanglement on (at least) one bipartition. Previous works on this question have provided evidence that the answer to this last question could be positive; to date, all the known nonlocal quantum states are distillable along some bipartition. In particular, Acin \cite{Toni} has shown that the asymptotic violation of the Mermin Bell inequalities, involving $N$ parties, always implies entanglement distillability. Later, this result was extended to other important classes of Bell inequalities: Ref. \cite{Lluis} addressed the case of all correlation Bell inequalities featuring two binary measurements per party \cite{WWZB}; Ref. \cite{Dani}, the case of a family of Bell inequalities for systems of arbitrary dimension introduced in Ref. \cite{cavalcanti}; Ref. \cite{Alejo}, the case of Bell inequalities based on multilinear contractions. Finally, it was shown that all bipartite entangled states display some hidden nonlocality \cite{masanes}.

Here we show however that quantum nonlocality does not imply distillability of entanglement in general. We present a family of 3-qubit quantum state which are fully bi-separable, i.e. separable along any bipartition of the parties. Hence no bipartite entanglement can be distilled between any groups of parties. We prove nevertheless analytically that a specific state in our family violates a simple tripartite Bell inequality presented in Ref \cite{sliwa}. Thus our state is fully bound entangled. Finally we discuss the relation between our work and the longstanding Peres conjecture \cite{peres99}, stating that bound entangled states always admit a local model, and can thus never violate a Bell inequality. Notably, our fully bound entangled state provides a counter-example to the strongest version of the Peres conjecture in the multipartite case.

\section{Family of 3-qubit biseparable states}

We consider a scenario featuring three distant parties (Alice, Bob and Charlie), sharing the 3-qubit quantum state

\begin{equation}
\rho=\sum_{j=1}^4 p_j \ket{\psi_j} \bra{\psi_j}.
\label{rhoabc}
\end{equation}
with $\sum_{j}p_j =1$, and \ba \label{param} \nonumber
\ket{\psi_1}&=&a_1 \ket{000} -b_1
(\ket{001}+\ket{010}+\ket{100})+c_1\ket{111} \\\nonumber
\ket{\psi_2}&=&-a_2 (\ket{001}-2\ket{010}+\ket{100}) \\ & & +b_2
(\ket{011}-2\ket{101}+\ket{110}) \\\nonumber \ket{\psi_3 }&=& a_3
(\ket{100}-\ket{001})  +b_3 (\ket{110}-\ket{011}) \\ \nonumber
\ket{\psi_4}&=&-a_4 \ket{000} +b_4
(\ket{011}+\ket{101}+\ket{110})+c_4\ket{111} \ea We choose the
parameters such that $p_2=p_3$ and
\begin{align}
\label{angles} a_1&=\sin\alpha\sin\beta\quad a_4=\cos\alpha\sin\gamma \nonumber\\
b_1&=\cos\beta/\sqrt 3 \quad b_4=\cos\gamma/\sqrt 3\nonumber\\
c_1&=\cos\alpha\sin\beta \quad c_4=\sin\alpha\sin\gamma \\
 a_2&=\cos\omega/\sqrt 6\quad a_3=\cos\omega/\sqrt 2 \nonumber\\
 b_2&=\sin\omega/\sqrt 6\quad b_3=\sin\omega/\sqrt 2. \nonumber
\end{align}
This ensures that $\braket{\psi_j}{\psi_k}=\delta_{jk}$, and that
the state $\rho$ is symmetric under any permutation of the
parties. Next we impose the relations $\rho_{000,011}=\rho_{001,010}$,
$\rho_{010,111}=\rho_{011,110}$, $\rho_{000,111}=\rho_{001,110}$, which ensure
that $\rho=\text{PT}_C(\rho)$ (where PT denotes the partial
transposition \cite{PPT} with respect to C). From the result of Ref.
\cite{kraus}, it then follows that $\rho$ is biseparable on the
bipartition $AB|C$. This leads to the following set of equations
\ba \label{eqs}
-a_4b_4p_4 &=&b_1^2p_1-2a_2^2p_2\nonumber\\
-b_1c_1p_1&=&-2b_2^2p_2+b_4^2p_4\\ \nonumber
-4a_2b_2p_2&=&a_1c_1p_1-a_4c_4p_4.\ea Note that since $\rho$ is
symmetric under permutation of the parties, it follows that $\rho$
is fully biseparable (i.e. biseparable along any bipartite cut) when conditions \eqref{eqs} hold. We
now solve the first two equations of \eqref{eqs}, together with
the normalization constraint $p_1+2p_2+p_4=1$, for the variables
$p_1$, $p_2$, and $p_4$. We get the following closed form \ba
\left(\begin{array}{c}
p_1\\
p_2\\
p_4\end{array}\right)
=M^{-1}\left(\begin{array}{c}
0\\
0\\
1\end{array}\right) =q\left(\begin{array}{c}
2a_4b_2^2b_4-2a_2^2b_4^2\\
-b_1^2b_4^2+a_4b_1b_4c_1\\
-2b_1^2b_2^2+2a_2^2b_1c_1
\end{array}\right)
\label{closed} \ea where $q=1/\det(M)$ is a normalization factor
and the matrix $M$ is given by, \ba M=\left(\begin{array}{ccc}
b_1^2 & -2a_2^2 & a_4b_4\\
b_1c_1 & -2b_2^2 & b_4^2\\
1 & 2 & 1\end{array}\right). \ea Then, substituting the above
solution for the weights $p_1$, $p_2$, and $p_4$ into the third
equation of \eqref{eqs}, we obtain the condition \ba
2a_2b_2A(\alpha,\beta,\gamma)+a_2^2B(\alpha,\beta,\gamma)+b_2^2C(\alpha,\beta,\gamma)=0
\label{ellipse} \ea where \ba \nonumber
A(\alpha,\beta,\gamma)&=& b_1b_4(a_4c_1-b_1b_4)  \\
B(\alpha,\beta,\gamma)&=& -c_1(a_1b_4^2+a_4b_1c_4) \\ \nonumber
C(\alpha,\beta,\gamma)&=&  a_4(a_1b_4c_1+b_1^2c_4) \ea Keeping in
mind the form of $a_2$ and $b_2$ (see \eqref{angles}), we see that
the solution is given by the intersection of two straight lines
crossing the origin (equation \eqref{ellipse}) and a circle
($a_2^2+b_2^2=1/6$), resulting in four different analytical
solutions for $\omega$ provided $A^2>BC$:
\ba \omega=\arctan{\frac{-A\pm\sqrt{A^2-BC}}{C}}\ea 
and the ones by adding the value of $\pi$. Thus for a choice of the 3 parameters,
$\alpha,\beta,\gamma$ \footnote{Note however that due to the
condition $A^2>BC$ not all triples $(\alpha,\beta,\gamma)$ lead to
a valid solution.}, the value of $\omega$ is given by the above
formulae (from which we are free to choose one), while the weights
$p_j$ are given by \eqref{closed}. Note that the positivity of the
weights $p_j$ is not guaranteed at this point. From
\eqref{closed}, we get two supplementary independent inequalities
of the form \ba \nonumber \tan\gamma &\ge& \frac{1}{\sqrt
3\cos\alpha}\frac{1}{\tan^2\omega}
\\  \tan\beta &\ge& \frac{1}{\sqrt 3\cos\alpha}\tan^2\omega, \ea
when the normalization factor $q=1/\det(M)$ is positive; when
$q<0$, both inequality signs must be reversed.

Thus we obtain a family of 3-qubit states, parametrized by the 3 parameters, $\alpha,\beta,\gamma$. Any state in this family is symmetric under any permutation of the parties, and fully biseparable, therefore undistillable along any bipartition. In the next section we show however that for judiciously chosen parameters, a state in this family can violate a Bell inequality (see Fig.~1). Indeed this implies that the state is bound entangled.

\begin{figure}[t]
  \includegraphics[width=0.8\columnwidth]{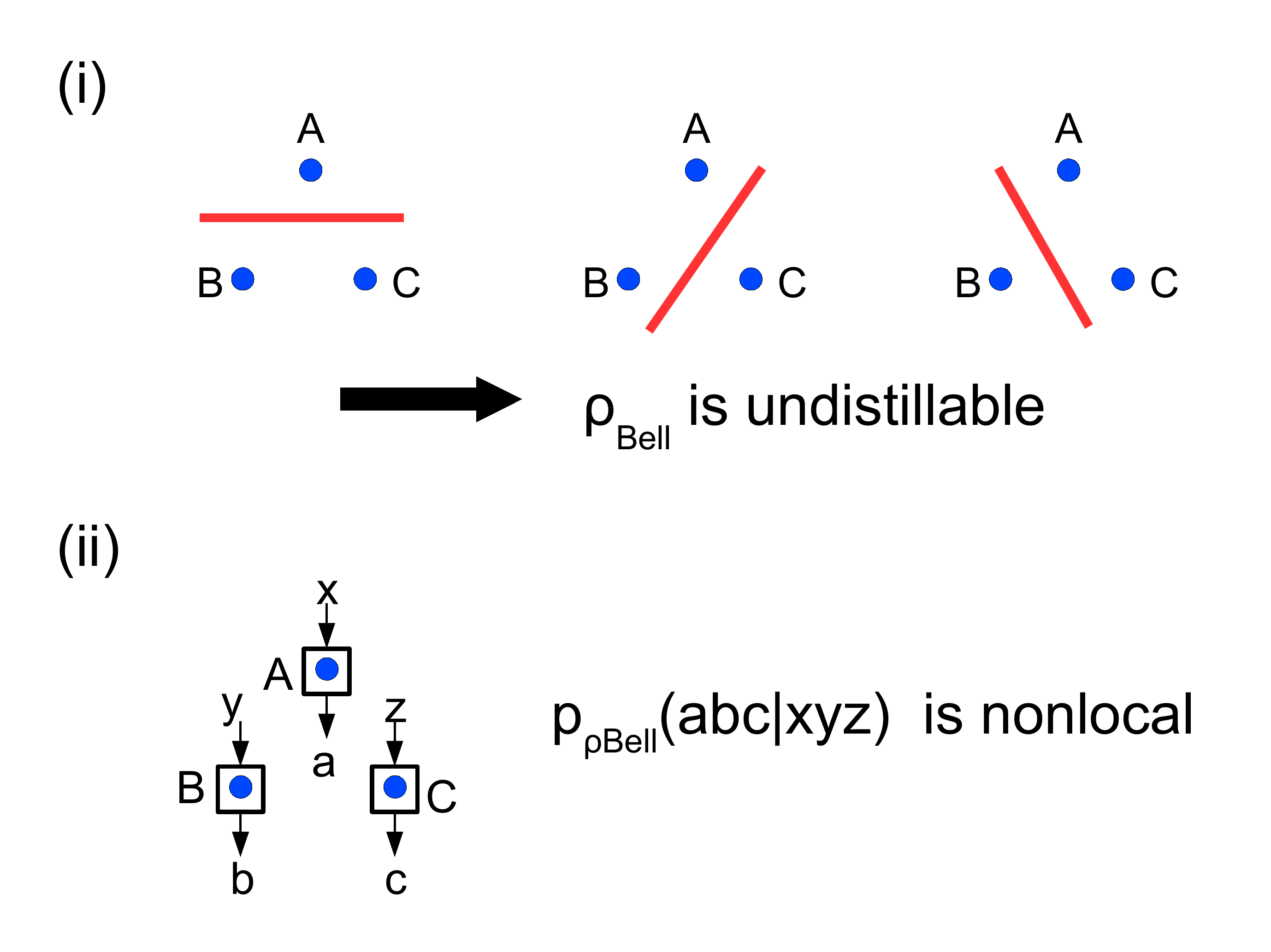}
  \caption{Properties of the 3-qubit state $\rho_{\text{Bell}}$. (i) Alice, Bob and Charlie share $\rho_{\text{Bell}}$ which is fully biseparable. Thus no pure bipartite entanglement can be distilled on any bipartition. (ii) It turns out however that $\rho_{\text{Bell}}$ can nevertheless violate a Bell inequality when judiciously chosen local measurements are performed. The joint probability distribution obtained in this way, i.e. $p_{\rho_{\text{Bell}}}(abc|xyz)$, is nonlocal. This shows that quantum nonlocality does not imply entanglement distillability.}
\label{fig}
\end{figure}

\section{Bell inequality violation}

We consider a Bell scenario where each of the three parties features two possible dichotomic measurements. We denote the outcome of measurement $j=1,2$ for Alice by $A_j$ ($B_j$ and $C_j$ for Bob and Charlie respectively). In Ref. \cite{sliwa}, all Bell inequalities have been obtained for this particular scenario. Here we shall focus on one of them, number 5 of \cite{sliwa}, which is given by the expression:
\ba\label{bell} S &=& \text{sym}[A_1 + A_1B_2 - A_2B_2 - A_1B_1C_1 \\\nonumber & &- A_2B_1C_1 + A_2B_2C_2]   \ea
where the notation $\text{sym}[X]$ means that the expression $X$ must be symmetrized over the parties, for instance $\text{sym}[A_1B_1]=A_1B_1+A_1C_1+B_1C_1$. The above polynomial should be understood as a sum of expectation values; for instance $A_1B_1$ means the expectation value of the product of the outcomes of Alice and Bob when they both perform the first measurement. All local probability distributions satisfy the inequality $S\leq 3$, which is therefore a Bell inequality.

We now show that by performing local measurements on a state of
the form (\ref{rhoabc},\ref{param}), it is possible to violate the
above Bell inequality, i.e. to get $S>3$. The local measurements
used by each party are of the simple form $A_j =
\cos{\theta_j}\sigma_z + \sin{\theta_j}\sigma_x$ (with $j=1,2$),
where $\sigma_x$ and $\sigma_z$ are Pauli matrices. Moreover, the
two  local measurements are the same for each party, i.e.
$A_j=B_j=C_j$. We could construct analytically a state $\rho_{\text{Bell}}$ and local measurements leading to a value of $S \approx 3.0069$, thus indicating a clear Bell violation. The parameters
characterizing this state $\rho_{\text{Bell}}$ are given by:
$\alpha=\pi/12$, $\beta=\pi/4$ and $\gamma=5\pi/12$, which entails
for the remaining parameters $\omega\approx0.5682$, $p_1\approx0.0636$,
$p_2=p_3\approx 0.2737$ and $p_4 \approx 0.3890$. The measurements are
given by $\theta_1=2\pi/9$ and $\theta_2=-4\pi/9$. More details
can be found in \footnote{A mathematica file containing all
details can be downloaded from http://www.atomki.hu/TheorPhys/Peres/Peres.nb.}. To the best of our knowledge, the state $\rho_{\text{Bell}}$ does not belong to any known class of bound entangled states; in particular it is not of the form of 'shift states' \cite{UPB} which are only symmetric under cyclic permutation of the parties.

Note that we could find numerically another bound entangled state of the form (\ref{rhoabc},\ref{param}) leading to a slightly higher violation of $S \approx 3.0187$ (see Appendix).

\section{Discussion}

We have presented a class of symmetric 3-qubit states which are fully biseparable and thus undistillable along any bipartition. Then we have shown that a particular state in this family, associated with judiciously chosen local measurements, violates a simple tripartite Bell inequality. Our result therefore demonstrate that entanglement distillability and nonlocality, two important forms of unseparability in quantum theory, are not equivalent.

Our result has also consequences for the Peres conjecture. This conjecture was initially formulated in the bipartite case, although Peres mentioned the multipartite extension in the conclusion of his original paper \cite{peres99}. Subsequently, two different extensions of the Peres conjecture to the multipartite case have been considered, which are based on different definitions of entanglement distillability (in the mulitpartite case). The first extension uses a weak definition of distilability, in which the parties are not allowed to group in order to distill bipartite entanglement. This version of the Peres conjecture has already been disproven by the results of Refs \cite{Dur}. The second extension of the conjecture, considered for instance by the authors of Refs \cite{RMP_horo,Alejo,WW}, uses the strongest definition of distillability, in which parties are allowed to group; this is the definition we used here. Our result thus disproves this second and stronger version of the Peres conjecture in the multipartite case, which had remained open up until now.

We have also investigated the link between entanglement distillability and nonlocality in the bipartite case. Using the techniques of Ref. \cite{seesaw}, we have performed extensive numerical research, considering Bell inequalities with up to six measurements per side \cite{BI} and quantum systems of dimension up to five, but could not find any example of a bipartite bound entangled state violating a Bell inequality. Thus the Peres conjecture remains open in the bipartite case.

Finally, our result could have interesting consequences in quantum information. On the one hand, entanglement distillability is generally associated to usefulness for processing quantum information, since most protocols use maximally entangled pure states. On the other hand, nonlocality is also associated to an advantage over classical models for certain tasks, such as communication complexity \cite{CC} and device-independent quantum key dstribution \cite{DI}. In this respect it would be interesting to get a better understanding of the Bell inequality that we used here, in particular to find whether a simple nonlocal game can be associated to it, and understand why bound entanglement provides an advantage.

\emph{Acknowledgements.} We thank Daniel Cavalcanti for discussions. The authors acknowledge financial support from the Hungarian National
Research Fund OTKA (PD101461) and from the UK EPSRC.

\section{Appendix}

Numerically, we found that the largest violation of the Bell
inequality \eqref{bell} using states of the form
(\ref{rhoabc},\ref{param}) was obtained for the following
parameters: $\alpha=0.1545$, $\beta=0.8460$ and $\gamma=4.4903$,
which entails for the remaining parameters $\omega\approx0.4808$,
$p_1\approx0.0338$, $p_2=p_3\approx0.2433$ and $p_4\approx0.4796$. The measurements
are given by $\theta_1=0.6897$ and $\theta_2=-1.2956$. This gives
a violation of $S \approx 3.0187$.

\end{document}